\documentclass[amsmath, aps, preprintnumbers, showpacs, twocolumn]{revtex4}
\usepackage{amsthm}
\usepackage{hyperref}
\usepackage{natbib}

\newcommand{\ra}{\rangle}
\newcommand{\la}{\langle}
\DeclareMathOperator{\diag}{diag}
\DeclareMathOperator{\Tr}{Tr}
\newtheorem{theorem}{Theorem}

\newtheorem{defi}{Definition}
\newtheorem{obs}{Observation}

\newcommand{\eprint}[2][]{\href{http://arxiv.org/abs/#2}{#2}}

\begin{document}
\title{Characterization of distillability of entanglement in terms of positive maps}
\author{Lieven Clarisse}
\email{lc181@york.ac.uk}
\affiliation{Dept. of Mathematics, University of York, Heslington, York, Y010 5DD, U.K.}

\begin{abstract}
A necessary and sufficient condition for one-distillability is formulated in terms of decomposable positive maps. As an application we provide insight into why all states violating the reduction criterion map are distillable and demonstrate how to construct such maps in a systematic way. We establish a connection between a number of existing results, which leads to an elementary proof for the characterization of distillability in terms of two-positive maps. 
\end{abstract}
\maketitle

\section{Introduction}
\emph{Distillation} is the process of converting with some finite probability a large number of mixed entangled states into a smaller number of maximally entangled pure states.
It relies on local (collective) manipulation of the particles together with classical communication between the parties (LOCC). It has become apparent that characterizing and quantifying distillation is of great importance in understanding the nature of entanglement from a physical point of view. Considerable effort has been devoted to characterizing distillable states, but even for bipartite systems the matter is not settled. In this paper we give a new necessary and sufficient condition for a state to be one-distillable, which includes several known criteria in a natural and simple way.

The distillation of pure states was completely analyzed in Ref.\ \cite{BBPS96}, in which it was shown that all entangled pure states can be reversibly distilled: in the limit of a large number of pairs one can extract the same number of singlet states as were needed to construct the state. Mixed state distillation proved to be much harder. The first papers were mainly concerned with qubit systems \cite{BBPSSW96, Gisin96}. Building upon these early results it was proven \cite{HHH97} that all entangled $2\otimes2$ mixed states can be distilled, but that distillation of mixed states requires inherently collective measurements \cite{LMP98, Kent98}. In Ref.\ \cite{HHH98} a necessary and sufficient condition was formulated for bipartite distillability:
\begin{theorem}[Horodecki et al.\ \cite{HHH98}]
\label{distilth}
A state $\rho$ is distillable if and only if there exist some two-dimensional projectors $P:{\cal H}^{\otimes n}_A \rightarrow {\cal H}^{\otimes n}_A $ and  $Q:{\cal H}^{\otimes n}_B \rightarrow {\cal H}^{\otimes n}_B $ and a number $n$, such that the state 
\begin{align}
\label{pqpq}
\rho'=(P\otimes Q)\rho^{\otimes n} (P\otimes Q)
\end{align}
is entangled.
\end{theorem}
As $\rho'$ lives in a $2\otimes 2$ space, a necessary and sufficient condition for distillability is that $\rho'$ has a negative partial transpose (NPT). The physical interpretation of this is clear: if we find such a two-dimensional subspace we can project upon it and distill the effectively two-qubit pair using known distillation protocols. The pure singlets can then be converted into maximally entangled states of the full space \cite{BBPS96}. A direct consequence of the theorem is that states with a positive partial transpose can never be distilled.
If the condition of the theorem is satisfied for a particular number $n$ then we call the state \emph{pseudo-$n$-copy distillable} or in short \emph{$n$ distillable}.
This result has proven to be extremely useful and is often used as a definition of distillability. It has recently been proven that for arbitrary $n$ there exist states that are $n$ distillable but not $(n-1)$ distillable \cite{Watrous03}.

It is conjectured that some NPT states cannot be distilled at all, and proving this is a central open problem \cite{DCLB99, DSSTT00}. The conjecture has the highly nontrivial consequence that bipartite distillable entanglement is nonadditive \cite{SST01}. Partial results are abundant and we would like to mention two results of great interest. Let us first fix some notation and recall the notion of positive map.

The set of bounded operators on a Hilbert space $\cal H$ is denoted by $B({\cal H})$. For $R \in B({\cal H})$, $R^T$ denotes the transpose with respect to  some given basis of $\cal H$; for $R \in B({\cal H}_A \otimes {\cal H}_B)$, $R^{T_A}$ and $R^{T_B}$ denote the partial transposes.
\begin{defi}
\emph{A \emph{positive map} $\Lambda:  B({\cal H}) \rightarrow  B({\cal H}) $ is a linear (not necessarily trace preserving) map between operators which preserves positivity. A $k$-\emph{positive map} is a positive map such that the induced map} 
\begin{align}
{\openone_k} \otimes \Lambda:B({\cal H}_{k}\otimes {\cal H})\rightarrow B({\cal H}_{k}\otimes {\cal H})
\end{align}
\emph{is positive. A \emph{completely positive map} (CP-map) is a map which is $k$ positive for all $k$ (or equivalently $d$ positive, with $d$ the dimension of ${\cal H}$).}
\end{defi}

It was shown \cite{HHH96} that if $\Lambda$ is a positive (but not necessarily CP) map then $\openone_k \otimes \Lambda$ is positive on separable states. Therefore positive maps can detect entanglement, the most famous example being the transpose. Another (weaker) map is given by $\Lambda_1(A)=\Tr(A)\openone-A$; this gives rise to the reduction criterion \cite{HH97, CAG97}. It was shown that all states violating the reduction criterion can be distilled. The second result we would like to mention is the following.
\begin{theorem}[DiVincenzo et al. \cite{DSSTT00}]
\label{divi}
Let $\rho$ be a state and  $S$ be the completely positive map defined by
\begin{align}
\rho=(\openone \otimes S)P_+ \hspace{1cm} \mbox{\emph{with}} \hspace{1cm} P_+=\frac{1}{d}\sum_{ij} |ii\ra\la jj|. \nonumber
\end{align}
Then $\rho$ is one-distillable if and only if $\Lambda=T\circ S$ is not two-positive (here $T$ denotes the transpose map).
\end{theorem}

As deciding whether all NPT states can be distilled currently seems too difficult, it is natural to ask if we can get stronger results in the other direction. In other words, what is the special role which the reduction criterion seems to play? Can we find other, perhaps stronger positive maps with the same property? In what follows we will answer these questions. In particular we will show how both of the above results follow naturally from the more general notion of \emph{distillability witness} and how to construct maps like that of the reduction criterion in a systematic way.

\section{Main results}
As separable states form a convex compact set, there exists an entanglement witness for each entangled state, i.e.\ a hyperplane which separates the entangled state from the set of separable states. Using the Jamio{\l}kowski isomorphism \cite{Jamiolkowski72} between operators and maps, this can be translated in terms of positive maps. We will pursue here the same line of reasoning for distillable states, and most of the proofs will be similar to the ones presented in Ref.\ \cite{HHH96}. Note that the distillation problem has been studied in connection with \emph{entanglement} witnesses \cite{KLC01}. Although this approach turned out to be very useful to obtain results about the activation properties of the state, it does not yield constructive tests for distillability since it depends on verifying whether a certain operator is an entanglement witness. For an excellent review on the characterization of convex sets and witnesses see Ref.\ \cite{BCHHKLS01}; see also Ref.\ \cite{SBL00}. 

The crux is that one-undistillable states also form a convex compact set. Indeed, from Theorem~\ref{distilth} and the linearity of the partial transpose, it follows that mixing one-undistillable states can never yield a distillable state \footnote{This does not contradict the conjecture of non convexity of distillable entanglement of bipartite states in \cite{SST01}, since we are working with a fixed number of pairs.}. What are the corresponding witnesses? To see this better, notice that from Theorem~\ref{distilth} it follows \cite{DCLB99} that distillability is equivalent to the existence of a Schmidt rank-2 state $|\psi\ra=c_1|a_1,b_1\ra+c_2|a_2,b_2\ra$,
with $\{ |a_1\ra, |a_2\ra\}$ two orthonormal vectors in ${\cal H}^{\otimes n}_A$ and $\{ |b_1\ra, |b_2\ra\}$ two orthonormal vectors in ${\cal H}^{\otimes n}_B$, and some $n$ such that 
\begin{align}
\la \psi | (\rho^{\otimes n})^{T_B}) |\psi\ra=\la \psi | (\rho^{T_B})^{\otimes n}) |\psi\ra<0.
\end{align}
To study the phenomenon of distillation it is sufficient to characterise the one-distillable states; $n$-distillable states $\rho$ can be characterised by looking at one copy of $\rho^{\otimes n}$. Thus we have the following.
\begin{obs}
For each one-distillable state $\sigma$ there exists an operator $D$ such that
$\Tr{D\sigma}<0$, but $\Tr{D\rho}\geq 0$ for all one-undistillable states $\rho$. The operator $D$ can always be chosen to be equal to $|\psi\ra\la\psi|^{T_B}$ for some Schmidt rank-2 vector $|\psi\ra$. 
\end{obs}
This theorem is compatible with the fact that PPT states cannot be distilled. Note that $D$ is decomposable by construction, i.e., can be written as $P+Q^{T_B}$ with $P,Q>0$, here $P=0$. For the following it is very interesting to consider slightly more general distillation witnesses $D$. Recall from Ref.\ \cite{TH00} the notion of generalised Schmidt number of density matrices: A bipartite density matrix $\rho$ has Schmidt number $k$ if
\begin{enumerate}
\item For any decomposition of $\rho$, $\{p_i\geq 0,|\psi_i\ra\}$ with $\rho=\sum_i p_i |\phi_i\ra\la\phi_i|$, at least one of the vectors $\{|\psi_i\ra\}$ has Schmidt rank at least $k$.
\item There exists a decomposition of $\rho$ with all vectors $\{|\psi_i\ra\}$ of Schmidt rank at most $k$.
\end{enumerate}
So separable states have Schmidt number 1; which is the same definition as for pure states.
The operators $D^{T_B}$ have thus Schmidt number 2. It is easy to see that in the above theorem we can allow all operators $D^{T_B}$ with Schmidt number 2.

The Jamio{\l}kowski isomorphism \cite{Jamiolkowski72} gives a one-to-one correspondence between linear maps $\Lambda:  B({\cal H}) \rightarrow  B({\cal H}) $ and operators $D \in B({\cal H} \otimes {\cal H})$, according to
\begin{align}
D&=d(\openone_d \otimes\Lambda)P_+, \\ 
\Lambda(X)&=\Tr_A{[D.(X^T\otimes\openone_d)]}. \label{amap}
\end{align}
In our case $D$ is a decomposable operator, therefore $\Lambda$ is a decomposable map ($\Lambda= T \circ \Lambda_1^{CP}+\Lambda_2^{CP}$, but here $\Lambda_2^{CP}=0$). What more can we say about the completely positive map $\Lambda_1^{CP}$? The general form \cite{Choi75, Kraus83} of such a map is given by
\begin{align}
\label{krauss}
\Lambda^{CP}(A)=\sum_i{V_i A V^\dagger_i},
\end{align}
where $V_i$ are arbitrary operators. Taking $D^{T_B}=|\psi\ra\la\psi|$, with $|\psi\ra=c_1|a_1,b_1\ra+c_2|a_2,b_2\ra$ a Schmidt rank-2 vector, we obtain after some algebra that the associated map is given by $\Lambda^{CP}(A)=V A V^\dagger$, with 
$V=c_1 |b_1\ra\la a_1|+c_2 |b_2\ra\la a_2|$.
Therefore for general $D^{T_B}$ with Schmidt number 2, the associated map satisfies (\ref{krauss}), with each $V_i$ an arbitrary operator of rank 2.
We call maps $\Lambda= T \circ \Lambda^{CP}= \tilde\Lambda^{CP}  \circ T$ defined in this way \emph{two-decomposable} \footnote{This differs from the term $k$ decomposability used in \cite{LMM03}, where it is used to denote maps that can be written as $ T \circ \Lambda_1^{k}+\Lambda_2^{k}$, with $\Lambda_i^{k}$ a $k$-positive map.}. 
With each map $\Lambda: B({\cal H}_n) \rightarrow B({\cal H}_m)$ there is associated an \emph{adjoint map} $\Lambda^\dagger: B({\cal H}_m) \rightarrow B({\cal H}_n)$ defined by $\Tr{(A\Lambda(B))}=\Tr{(\Lambda^\dagger(A)B)}$ for all $A$ and $B$. It is easy to show that 
if $\Lambda$ is two-decomposable then the adjoint map $\Lambda^\dagger$ is two-decomposable. 

\begin{theorem}[Main Theorem]
\label{maintheorem}
A state $\rho$ is one-undistil\-lable if and only if
\begin{align}
(\openone\otimes \Lambda)(\rho)\geq 0
\end{align}
for all two-decomposable maps $\Lambda$.
\end{theorem}
\begin{proof}
Suppose $\rho$ is one-distillable, so that there exists a $D=|\psi\ra\la\psi|^{T_B}$, with $|\psi\ra$ a Schmidt rank-2 vector such that $\Tr(D\rho)<0$. Using the associated map (\ref{amap}) this can be written as
\begin{align}
\Tr[(\openone \otimes\Lambda)(P_+)\rho]=\Tr[(\openone \otimes\Lambda^\dagger)(\rho)P_+]<0,
\end{align}
and since $P_+$ is positive it follows that $\openone \otimes\Lambda^\dagger(\rho)\not\geq 0$, i.e.\ we have found a two-decomposable map that detects the state. 

To prove the converse, let $\Lambda=\Lambda^{CP}\circ T$ be a two-decomposable map such that $\openone \otimes\Lambda(\rho)$ has a negative eigenvalue. Denoting by $|\phi\ra$ the corresponding eigenvector, we get
\begin{align}
\label{keyeq}
\la\phi|\openone\otimes\Lambda^{CP}(\rho^{T_B}) |\phi\ra=\Tr[(\openone \otimes\Lambda^{\dagger})(|\phi\ra\la\phi|)\rho^{T_B}] <0.
\end{align}
It is sufficient to consider $\Lambda^{CP}(A)=V A V^\dagger$, with $V$ a rank-2 operator, therefore $\openone \otimes \Lambda^\dagger(|\phi\ra\la\phi|)= |\phi'\ra\la\phi'|$, with $|\phi'\ra$ Schmidt rank 2. Thus $\Tr( \rho^{T_B}|\phi'\ra\la\phi'|)<0$, and $\rho$ is one-distillable.
\end{proof}

Note that given a two-decomposable map $\Lambda$, the above result implies that undistillable states must satisfy both $(\openone\otimes \Lambda)(\rho)\geq 0$ and the dual criterion ($\Lambda\otimes \openone )(\rho)\geq 0$.

\section{Illustrations}
To illustrate the power of the above formulation, we will rederive two known results. 
The first is that all states violating the reduction criterion can be distilled. The original proof \cite{HH97} relied on a series of protocols: filtering, twirling and distillation of isotropic states. Let us construct the operator $D^{T_B}$ associated with the Schmidt rank-2 vectors $|\psi_{ij}\ra=|ij\ra-|ji\ra$, $i\neq j$. If we add the resulting operators we get
\begin{align}
D^{T_B}=\sum_{i<j} |\psi_{ij}\ra\la\psi_{ij}|=\sum_{i\neq j}{(|ij\ra\la ij|-|ij\ra\la ji|)}=\openone-V,
\end{align}
with $V=\sum_{i,j}|ij\ra\la ji|$ the flip operator, and thus $D=\openone-dP_+$. The associated map is given by $\Lambda_1(A)=\Tr(A)\openone-A$. This is just the map used by the reduction criterion for entanglement. Thus if a state $\rho$ satisfies $\openone\otimes \rho_B-\rho \not \geq 0$ or $\rho_A\otimes\openone -\rho\not \geq 0$ then it is one-distillable. In this way we have also proven that the map is decomposable, and using the results of the previous section we can obtain the explicit Kraus form (\ref{krauss}) for the map. 

The second result concerns the formulation of distillation in terms of two-positive maps (Theorem~\ref{divi}). We will provide an elegant proof of this, making use of the following theorem, which is implicit in Ref.\ \cite{TH00}: \emph{A map is k-positive if and only if the corresponding operator is positive on states with Schmidt number $k$ or less} \footnote{This is a generalisation and unification of the theorems by Choi \cite{Choi75} and Jamio{\l}kowski \cite{Jamiolkowski72}.}. Now, for undistillable states $\rho$ we have that 
\begin{align}
\Tr{(\rho D)}=\Tr{(\rho^{T_B} D^{T_B})}>0,
\end{align}
for all $D^{T_B}$ with Schmidt number 2. From the previously mentioned theorem we deduce that the map associated with $\rho^{T_B}$ is two-positive for undistillable states. In other words, let $S$ be the completely positive map defined by
\begin{align}
\rho=(\openone \otimes S)P_+.
\end{align}
and define the positive map $ \Lambda=T\circ S$; then $\rho$ is 1-undistillable if and only if $\Lambda$ is two-positive.

As illustrated with the reduction map, distillation witnesses can be obtained by combining Schmidt rank-2 vectors. Here we give some more examples.

1. Taking our vectors $|ij\ra+|ji\ra$ we obtain
\begin{align}
D^{T_B}=\sum_{i\neq j}{(|ij\ra\la ij|+|ij\ra\la ji|)}=\openone+V-2Z,
\end{align}
with $Z=\sum_i|ii\ra\la ii|$. Therefore $D=\openone+dP_+ - 2Z$ and the associated map is given by $\Lambda_2(A)=\Tr A\openone+A-2\diag{A}$. The map $diag$ maps all off-diagonal elements to zero and leaves the diagonal itself invariant.

2. For the Schmidt rank-2 vectors of the form $|ii\ra+|jj\ra$, we get:
\begin{align}
D^{T_B}=\sum_{i\neq j}{(|ii\ra\la ii|+|ii\ra\la jj|)}=dP_+ +(d-2)Z,
\end{align}
and thus $D=V+ (d-2)Z$; the associated map is given by 
$\Lambda_3(A)=A^T+(d-2)\diag{A}$. 

3. If we take Schmidt rank-2 vectors of the form $|ii\ra-|jj\ra$, we get:
\begin{align}
D^{T_B}=\sum_{i\neq j}{(|ii\ra\la ii|-|ii\ra\la jj|)}=-dP_+ +dZ,
\end{align}
and thus $D=-V+ dZ$; the associated map is given by 
$\Lambda_4(A)=-A^T+d\diag{A}$.

In fact, \emph{every} operator with Schmidt number 2 gives us a strong distillation witness:

4. In \cite{TH00} it is proven that the special isotropic state
\begin{align}
D^{T_B}=(d-2)\openone+(2d-1)dP_+
\end{align}
allows a Schmidt rank-2 decomposition. We find $D=(d-2)\openone+(2d-1)V$ and for the corresponding map $\Lambda_5(A)=(d-2)\Tr A \openone + (2d-1)A^T$.

\section{Discussion and Conclusion}
As is well known in the theory of characterization of convex sets with aid of witnesses (see for instance Ref.\ \cite{Alb01}), the dual formulation in terms of positive maps is much stronger. That is, the map detects \emph{more} states. Let us redo the second part of the main theorem in a slightly different way to see this more clearly. So suppose $\openone \otimes \Lambda(\rho) \not\geq 0$; then we can rewrite equation \ref{keyeq} as 
\begin{align}
\Tr[(A\otimes \openone) D (A^{\dagger}\otimes \openone) \rho] <0,
\end{align}
where we have substituted $|\phi\ra\la\phi |=(A\otimes \openone) P_+ (A^{\dagger}\otimes \openone)$. In other words the map $\Lambda$ corresponds to the class of witnesses $(A\otimes \openone) D (A^{\dagger}\otimes \openone)$, for arbitrary $A$. 

It also implies that the criteria $\openone \otimes \Lambda(\rho) \geq 0$ and $\Lambda(\rho)  \otimes \openone\geq 0$ are insensitive to local transformations by one of the parties. Indeed, suppose Alice performs a general measurement, with measurement operators $A_1$ and $A_2$ satisfying $A^\dagger_1A_1 +A^\dagger_2A_2=\openone$ (a so called filtering operation). Then the state $\rho$ will be transformed into $\rho_i=(A_i\otimes \openone) \rho (A^{\dagger}_i\otimes \openone)/p_i$. It follows that, if the original state did not violate the criteria, then the transformed state doesn't either. The map corresponds to the operator witness, together with all possible local filtering operations by one of the parties.

We have shown how to construct maps that detect distillability, applicable in arbitrary dimensions, which  can be easily evaluated on states. There is however a catch.  As can be seen from the reconstruction of the reduction criterion, the witness from which the map is derived, is a convex combination of \emph{many} witnesses. So it is possible that one of those witnesses (and the associated map) detects a state, while the sum doesn't. In other words, the map could be weaker than expected at first sight. The sum of the negative eigenvalues of $D$, or the entanglement of $D^{T_B}$ can be seen as a measure for the strength of a certain map.

The characterization of the separability problem in terms of positive maps faces the problem that a complete characterization of positive maps is unknown. In contrast, the cone of two-decomposable maps is complete characterized (by definition). Moreover as shown above, application of the map, replaces the distillation witness (in terms of the operator), with a whole class of distillation witnesses. In concreto, the minimization of $\la \psi | \rho^{T_B} |\psi\ra$, with $|\psi\ra$ Schmidt rank 2, can be replaced with a minimization of $\openone\otimes \Lambda(\rho)$ with $\Lambda$ the two-decomposable map corresponding to all \emph{maximally entangled} Schmidt rank-2 $|\psi\ra$.

In the context of separability, the formulation in terms of positive maps is not the only way to tackle the problem. In particular convex optimization techniques (see for instance Ref.\ \cite{EHGC04}) have been successfully applied. In a forthcoming paper we will discuss such techniques in the context of the distillability problem.

The basis requirement for a distillation witness $D$ to detect a two-distillable state $\rho$ which is not one-distillable is that $D^{T_B}$ must be entangled with respect to the first and second pair. Indeed suppose $D^{T_B}=|\psi\ra\la\psi|$ with $|\psi\ra=|\psi_1\ra\otimes |\psi_2\ra$ separable, then $\Tr(D^{T_B}\rho^{\otimes 2T_B})=\Tr(|\psi_1\ra\la\psi_1| \rho)\Tr(|\psi_2\ra\la\psi_2| \rho)>0$, since the vectors $|\psi_i\ra$ have at most Schmidt rank 2. Consider the witness for the reduction criterion on two pairs: $D^{T_B}=\openone -V= \openone_1 \otimes \openone_2 - V_1 \otimes V_2$. But from Ref.\ \cite{Alb01} $V=P_S-P_A$ and $\openone=P_S+P_A$ with $P_S$ and $P_A$  projection operators onto the symmetric and antisymmetric subspaces respectively. Substitution yields $D^{T_B}= P_{S1} \otimes P_{A2} + P_{A1}\otimes P_{S2}$  so that $D^{T_B}$ is separable with respect to the different pairs. Note that this property is not adhered by the map since $(A\otimes \openone) D^{T_B} (A^{\dagger}\otimes \openone)$ could be entangled with respect to the two pairs, even if $D$ itself isn't. So it is not at all obvious that collective application of the reduction criterion (for instance on $\rho ^{\otimes 2}$) does not yield a stronger criterion. It was proven in Ref.\ \cite{HH97} that it is not the case: they proved that if $\openone \otimes \Lambda_1 (\rho^{\otimes 2})\not\geq 0$ then $\openone \otimes \Lambda_1 (\rho)\not\geq 0$. It is readily verified that all introduced maps share this property. In the case of the reduction criterion, the reverse is also true. This can be proved as follows. From Ref.\ \cite{HH97} $\openone_A \otimes  \Lambda_B (\rho_1 \otimes \rho_2)= \openone_{A1} \otimes  \rho^B_1 \otimes \openone_{A2} \otimes  \Lambda_{B2} ( \rho_2) + \openone_{A1} \otimes  \Lambda_{B1} ( \rho_1) \otimes \rho_2$. Now suppose $\openone \otimes  \Lambda  (\rho_i)$ is not a positive operator, so that there exist a vector $|\psi\ra$ with negative expectation value. From the above expression follows that  $\la\Psi|\openone_A \otimes  \Lambda_B (\rho_1 \otimes \rho_2)|\Psi\ra<0$, with $|\Psi\ra=|\psi_1\ra \otimes |\psi_2\ra$. So applying the reduction criterion to one or more pairs is completely equivalent (this also applies for $\Lambda_4$).

In conclusion, we have clarified the role which the reduction criterion plays in the story of distillation: it is just an example of a two-decomposable positive map. The formulation of the distillation problem in terms of those positive maps gives rise to a class of strong criteria, that are more powerful than any other known criteria or reformulation of the problem.
It is our hope that the presented results will shed some light on the question of the convexity of the whole set of undistillable states. 

\begin{acknowledgments}
The author would like to thank A. Sudbery for interesting discussions and comments on this paper.
\end{acknowledgments}

%\bibliographystyle{apsrevl}
%\bibliography{ent}

\begin{thebibliography}{24}
\expandafter\ifx\csname natexlab\endcsname\relax\def\natexlab#1{#1}\fi
\expandafter\ifx\csname bibnamefont\endcsname\relax
  \def\bibnamefont#1{#1}\fi
\expandafter\ifx\csname bibfnamefont\endcsname\relax
  \def\bibfnamefont#1{#1}\fi
\expandafter\ifx\csname citenamefont\endcsname\relax
  \def\citenamefont#1{#1}\fi
\expandafter\ifx\csname url\endcsname\relax
  \def\url#1{\texttt{#1}}\fi
\expandafter\ifx\csname urlprefix\endcsname\relax\def\urlprefix{URL }\fi
\providecommand{\bibinfo}[2]{#2}
\providecommand{\eprint}[2][]{\url{#2}}

\bibitem[{\citenamefont{Bennett
  et~al.}(1996{\natexlab{a}})\citenamefont{Bennett, Bernstein, Popescu, and
  Schumacher}}]{BBPS96}
\bibinfo{author}{\bibfnamefont{C.~H.} \bibnamefont{Bennett}},
  \bibinfo{author}{\bibfnamefont{H.~J.} \bibnamefont{Bernstein}},
  \bibinfo{author}{\bibfnamefont{S.}~\bibnamefont{Popescu}}, \bibnamefont{and}
  \bibinfo{author}{\bibfnamefont{B.}~\bibnamefont{Schumacher}},
  \emph{\bibinfo{title}{Concentrating partial entanglement by local
  operations}}, \bibinfo{journal}{Physcial Review A}
  \textbf{\bibinfo{volume}{53}}, \bibinfo{pages}{2046}
  (\bibinfo{year}{1996}{\natexlab{a}}), \eprint{quant-ph/9511030}.

\bibitem[{\citenamefont{Bennett
  et~al.}(1996{\natexlab{b}})\citenamefont{Bennett, Brassard, Popescu,
  Schumacher, Smolin, and Wootters}}]{BBPSSW96}
\bibinfo{author}{\bibfnamefont{C.~H.} \bibnamefont{Bennett}},
  \bibinfo{author}{\bibfnamefont{G.}~\bibnamefont{Brassard}},
  \bibinfo{author}{\bibfnamefont{S.}~\bibnamefont{Popescu}},
  \bibinfo{author}{\bibfnamefont{B.}~\bibnamefont{Schumacher}},
  \bibinfo{author}{\bibfnamefont{J.~A.} \bibnamefont{Smolin}},
  \bibnamefont{and} \bibinfo{author}{\bibfnamefont{W.~K.}
  \bibnamefont{Wootters}}, \emph{\bibinfo{title}{Purification of noisy
  entanglement and faithful teleportation via noisy channels}},
  \bibinfo{journal}{Physical Review Letters} \textbf{\bibinfo{volume}{76}},
  \bibinfo{pages}{722} (\bibinfo{year}{1996}{\natexlab{b}}),
  \eprint{quant-ph/9511027}.

\bibitem[{\citenamefont{Gisin}(1996)}]{Gisin96}
\bibinfo{author}{\bibfnamefont{N.}~\bibnamefont{Gisin}},
  \emph{\bibinfo{title}{Hidden quantum nonlocality revealed by local filters}},
  \bibinfo{journal}{Physics Letters A} \textbf{\bibinfo{volume}{210}},
  \bibinfo{pages}{151} (\bibinfo{year}{1996}).

\bibitem[{\citenamefont{Horodecki et~al.}(1997)\citenamefont{Horodecki,
  Horodecki, and Horodecki}}]{HHH97}
\bibinfo{author}{\bibfnamefont{M.}~\bibnamefont{Horodecki}},
  \bibinfo{author}{\bibfnamefont{P.}~\bibnamefont{Horodecki}},
  \bibnamefont{and}
  \bibinfo{author}{\bibfnamefont{R.}~\bibnamefont{Horodecki}},
  \emph{\bibinfo{title}{Inseparable two spin-{$\frac{1}{2}$} density matrices
  can be distilled to a singlet form}}, \bibinfo{journal}{Physical Review
  Letters} \textbf{\bibinfo{volume}{78}}, \bibinfo{pages}{574}
  (\bibinfo{year}{1997}), \eprint{quant-ph/9607009}.

\bibitem[{\citenamefont{Linden et~al.}(1998)\citenamefont{Linden, Massar, and
  Popescu}}]{LMP98}
\bibinfo{author}{\bibfnamefont{N.}~\bibnamefont{Linden}},
  \bibinfo{author}{\bibfnamefont{S.}~\bibnamefont{Massar}}, \bibnamefont{and}
  \bibinfo{author}{\bibfnamefont{S.}~\bibnamefont{Popescu}},
  \emph{\bibinfo{title}{Purifying noisy entanglement requires collective
  measurements}}, \bibinfo{journal}{Physical Review Letters}
  \textbf{\bibinfo{volume}{81}}, \bibinfo{pages}{3279} (\bibinfo{year}{1998}),
  \eprint{quant-ph/9805001}.

\bibitem[{\citenamefont{Kent}(1998)}]{Kent98}
\bibinfo{author}{\bibfnamefont{A.}~\bibnamefont{Kent}},
  \emph{\bibinfo{title}{Entangled mixed states and local purification}},
  \bibinfo{journal}{Physical Review Letters} \textbf{\bibinfo{volume}{81}},
  \bibinfo{pages}{2839} (\bibinfo{year}{1998}), \eprint{quant-ph/9805088}.

\bibitem[{\citenamefont{Horodecki et~al.}(1998)\citenamefont{Horodecki,
  Horodecki, and Horodecki}}]{HHH98}
\bibinfo{author}{\bibfnamefont{M.}~\bibnamefont{Horodecki}},
  \bibinfo{author}{\bibfnamefont{P.}~\bibnamefont{Horodecki}},
  \bibnamefont{and}
  \bibinfo{author}{\bibfnamefont{R.}~\bibnamefont{Horodecki}},
  \emph{\bibinfo{title}{Mixed-state entanglement and distillation: is there a
  {`}bound{'} entanglement in nature?}}, \bibinfo{journal}{Physical Review
  Letters} \textbf{\bibinfo{volume}{80}}, \bibinfo{pages}{5239}
  (\bibinfo{year}{1998}), \eprint{quant-ph/9801069}.

\bibitem[{\citenamefont{Watrous}(2004)}]{Watrous03}
\bibinfo{author}{\bibfnamefont{J.}~\bibnamefont{Watrous}},
  \emph{\bibinfo{title}{On the number of copies required for entanglement
  distillation}}, \bibinfo{journal}{Physical Review Letters}
  \textbf{\bibinfo{volume}{93}}, \bibinfo{pages}{010502}
  (\bibinfo{year}{2004}), \eprint{quant-ph/0312123}.

\bibitem[{\citenamefont{D{\"u}r et~al.}(2000)\citenamefont{D{\"u}r, Cirac,
  Lewenstein, and Bru{\ss}}}]{DCLB99}
\bibinfo{author}{\bibfnamefont{W.}~\bibnamefont{D{\"u}r}},
  \bibinfo{author}{\bibfnamefont{J.~I.} \bibnamefont{Cirac}},
  \bibinfo{author}{\bibfnamefont{M.}~\bibnamefont{Lewenstein}},
  \bibnamefont{and} \bibinfo{author}{\bibfnamefont{D.}~\bibnamefont{Bru{\ss}}},
  \emph{\bibinfo{title}{Distillability and partial transpostion in bipartite
  systems}}, \bibinfo{journal}{Physical Review A}
  \textbf{\bibinfo{volume}{61}}, \bibinfo{pages}{062313}
  (\bibinfo{year}{2000}), \eprint{quant-ph/9910022}.

\bibitem[{\citenamefont{Di{V}incenzo et~al.}(2000)\citenamefont{Di{V}incenzo,
  Shor, Smolin, Terhal, and Thapliyal}}]{DSSTT00}
\bibinfo{author}{\bibfnamefont{D.~P.} \bibnamefont{Di{V}incenzo}},
  \bibinfo{author}{\bibfnamefont{P.~W.} \bibnamefont{Shor}},
  \bibinfo{author}{\bibfnamefont{J.~A.} \bibnamefont{Smolin}},
  \bibinfo{author}{\bibfnamefont{B.~M.} \bibnamefont{Terhal}},
  \bibnamefont{and} \bibinfo{author}{\bibfnamefont{A.~V.}
  \bibnamefont{Thapliyal}}, \emph{\bibinfo{title}{Evidence for bound entangled
  states with negative partial transpose}}, \bibinfo{journal}{Physical Review
  A} \textbf{\bibinfo{volume}{61}}, \bibinfo{pages}{062312}
  (\bibinfo{year}{2000}), \eprint{quant-ph/9910026}.

\bibitem[{\citenamefont{Shor et~al.}(2001)\citenamefont{Shor, Smolin, and
  Terhal}}]{SST01}
\bibinfo{author}{\bibfnamefont{P.~W.} \bibnamefont{Shor}},
  \bibinfo{author}{\bibfnamefont{J.~A.} \bibnamefont{Smolin}},
  \bibnamefont{and} \bibinfo{author}{\bibfnamefont{B.~M.}
  \bibnamefont{Terhal}}, \emph{\bibinfo{title}{Nonadditivity of bipartite
  distillable entanglement follows from conjecture on bound entangled {W}erner
  states}}, \bibinfo{journal}{Physical Review Letters}
  \textbf{\bibinfo{volume}{86}}, \bibinfo{pages}{2681} (\bibinfo{year}{2001}),
  \eprint{quant-ph/0010054}.

\bibitem[{\citenamefont{Horodecki et~al.}(1996)\citenamefont{Horodecki,
  Horodecki, and Horodecki}}]{HHH96}
\bibinfo{author}{\bibfnamefont{M.}~\bibnamefont{Horodecki}},
  \bibinfo{author}{\bibfnamefont{P.}~\bibnamefont{Horodecki}},
  \bibnamefont{and}
  \bibinfo{author}{\bibfnamefont{R.}~\bibnamefont{Horodecki}},
  \emph{\bibinfo{title}{Separability of mixed states: necessary and sufficient
  conditions}}, \bibinfo{journal}{Physics Letters A}
  \textbf{\bibinfo{volume}{223}}, \bibinfo{pages}{1} (\bibinfo{year}{1996}),
  \eprint{quant-ph/9605038}.

\bibitem[{\citenamefont{Horodecki and Horodecki}(1999)}]{HH97}
\bibinfo{author}{\bibfnamefont{M.}~\bibnamefont{Horodecki}} \bibnamefont{and}
  \bibinfo{author}{\bibfnamefont{P.}~\bibnamefont{Horodecki}},
  \emph{\bibinfo{title}{Reduction criterion of separability and limits for a
  class of distillation protocols}}, \bibinfo{journal}{Physical Review A}
  \textbf{\bibinfo{volume}{59}}, \bibinfo{pages}{4206} (\bibinfo{year}{1999}),
  \eprint{quant-ph/9708015}.

\bibitem[{\citenamefont{Cerf et~al.}(1997)\citenamefont{Cerf, Adami, and
  Gingrich}}]{CAG97}
\bibinfo{author}{\bibfnamefont{N.~J.} \bibnamefont{Cerf}},
  \bibinfo{author}{\bibfnamefont{C.}~\bibnamefont{Adami}}, \bibnamefont{and}
  \bibinfo{author}{\bibfnamefont{R.~M.} \bibnamefont{Gingrich}},
  \emph{\bibinfo{title}{Reduction criterion for separability}},
  \bibinfo{journal}{Physical Review A} \textbf{\bibinfo{volume}{60}},
  \bibinfo{pages}{898} (\bibinfo{year}{1997}), \eprint{quant-ph/9710001}.

\bibitem[{\citenamefont{Jamio{\l}kowski}(1972)}]{Jamiolkowski72}
\bibinfo{author}{\bibfnamefont{A.}~\bibnamefont{Jamio{\l}kowski}},
  \emph{\bibinfo{title}{Linear transformations which preserve trace and
  positive semidefiteness of operators}}, \bibinfo{journal}{Reports on
  Mathematical Physics} \textbf{\bibinfo{volume}{3}}, \bibinfo{pages}{275}
  (\bibinfo{year}{1972}).

\bibitem[{\citenamefont{Kraus et~al.}(2002)\citenamefont{Kraus, Lewenstein, and
  Cirac}}]{KLC01}
\bibinfo{author}{\bibfnamefont{B.}~\bibnamefont{Kraus}},
  \bibinfo{author}{\bibfnamefont{M.}~\bibnamefont{Lewenstein}},
  \bibnamefont{and} \bibinfo{author}{\bibfnamefont{J.~I.} \bibnamefont{Cirac}},
  \emph{\bibinfo{title}{Characterization of distillable and activable states
  using entanglement witnesses}}, \bibinfo{journal}{Physical Review A}
  \textbf{\bibinfo{volume}{65}}, \bibinfo{pages}{042327}
  (\bibinfo{year}{2002}), \eprint{quant-ph/0110174}.

\bibitem[{\citenamefont{Bru{\ss} et~al.}(2002)\citenamefont{Bru{\ss}, Cirac,
  Horodecki, Hulpke, Kraus, Lewenstein, and Florian}}]{BCHHKLS01}
\bibinfo{author}{\bibfnamefont{D.}~\bibnamefont{Bru{\ss}}},
  \bibinfo{author}{\bibfnamefont{J.~I.} \bibnamefont{Cirac}},
  \bibinfo{author}{\bibfnamefont{P.}~\bibnamefont{Horodecki}},
  \bibinfo{author}{\bibfnamefont{F.}~\bibnamefont{Hulpke}},
  \bibinfo{author}{\bibfnamefont{B.}~\bibnamefont{Kraus}},
  \bibinfo{author}{\bibfnamefont{M.}~\bibnamefont{Lewenstein}},
  \bibnamefont{and} \bibinfo{author}{\bibfnamefont{A.}~\bibnamefont{Florian}},
  \emph{\bibinfo{title}{Reflections upon separability and distillability}},
  \bibinfo{journal}{Journal of Modern Optics} \textbf{\bibinfo{volume}{49}},
  \bibinfo{pages}{1399} (\bibinfo{year}{2002}), \eprint{quant-ph/0110081}.

\bibitem[{\citenamefont{Sanpera et~al.}(2001)\citenamefont{Sanpera, Bru{\ss},
  and Lewenstein}}]{SBL00}
\bibinfo{author}{\bibfnamefont{A.}~\bibnamefont{Sanpera}},
  \bibinfo{author}{\bibfnamefont{D.}~\bibnamefont{Bru{\ss}}}, \bibnamefont{and}
  \bibinfo{author}{\bibfnamefont{M.}~\bibnamefont{Lewenstein}},
  \emph{\bibinfo{title}{Schmidt number witnesses and bound entanglement}},
  \bibinfo{journal}{Physical Review A} \textbf{\bibinfo{volume}{63}},
  \bibinfo{pages}{050301(R)} (\bibinfo{year}{2001}), \eprint{quant-ph/0009109}.

\bibitem[{\citenamefont{Terhal and Horodecki}(2000)}]{TH00}
\bibinfo{author}{\bibfnamefont{B.~M.} \bibnamefont{Terhal}} \bibnamefont{and}
  \bibinfo{author}{\bibfnamefont{P.}~\bibnamefont{Horodecki}},
  \emph{\bibinfo{title}{A {S}chmidt number for density matrices}},
  \bibinfo{journal}{Physical Review A} \textbf{\bibinfo{volume}{61}},
  \bibinfo{pages}{040301(R)} (\bibinfo{year}{2000}), \eprint{quant-ph/9911117}.

\bibitem[{\citenamefont{Choi}(1975)}]{Choi75}
\bibinfo{author}{\bibfnamefont{M.-D.} \bibnamefont{Choi}},
  \emph{\bibinfo{title}{Completely positive linear maps on complex matrices}},
  \bibinfo{journal}{Linear Algebra and its Applications}
  \textbf{\bibinfo{volume}{10}}, \bibinfo{pages}{285} (\bibinfo{year}{1975}).

\bibitem[{\citenamefont{Kraus}(1983)}]{Kraus83}
\bibinfo{author}{\bibfnamefont{K.}~\bibnamefont{Kraus}},
  \emph{\bibinfo{title}{States, Effects and Operations}}, vol.
  \bibinfo{volume}{190} of \emph{\bibinfo{series}{Lecture Notes in Physics}}
  (\bibinfo{publisher}{Springer}, \bibinfo{address}{Berlin, Heidelberg},
  \bibinfo{year}{1983}).

\bibitem[{\citenamefont{Alber et~al.}(2001)\citenamefont{Alber, Beth,
  Horodecki, Horodecki, Horodecki, R{{\"o}tteler}, Weinfurter, Werner, and
  Zeilinger}}]{Alb01}
\bibinfo{author}{\bibfnamefont{G.}~\bibnamefont{Alber}},
  \bibinfo{author}{\bibfnamefont{T.}~\bibnamefont{Beth}},
  \bibinfo{author}{\bibfnamefont{M.}~\bibnamefont{Horodecki}},
  \bibinfo{author}{\bibfnamefont{P.}~\bibnamefont{Horodecki}},
  \bibinfo{author}{\bibfnamefont{R.}~\bibnamefont{Horodecki}},
  \bibinfo{author}{\bibfnamefont{M.}~\bibnamefont{R{{\"o}tteler}}},
  \bibinfo{author}{\bibfnamefont{H.}~\bibnamefont{Weinfurter}},
  \bibinfo{author}{\bibfnamefont{R.}~\bibnamefont{Werner}}, \bibnamefont{and}
  \bibinfo{author}{\bibfnamefont{A.}~\bibnamefont{Zeilinger}},
  \emph{\bibinfo{title}{Quantum Information: An Introduction to Basic
  Theoretical Concepts and Experiments}}, vol. \bibinfo{volume}{173} of
  \emph{\bibinfo{series}{Springer Tracts in Modern Physics}}
  (\bibinfo{publisher}{Springer}, \bibinfo{year}{2001}), ISBN
  \bibinfo{isbn}{3540416668}.

\bibitem[{\citenamefont{Eisert et~al.}(2005)\citenamefont{Eisert, Hyllus,
  G{\"u}hne, and Curty}}]{EHGC04}
\bibinfo{author}{\bibfnamefont{J.}~\bibnamefont{Eisert}},
  \bibinfo{author}{\bibfnamefont{P.}~\bibnamefont{Hyllus}},
  \bibinfo{author}{\bibfnamefont{O.}~\bibnamefont{G{\"u}hne}},
  \bibnamefont{and} \bibinfo{author}{\bibfnamefont{M.}~\bibnamefont{Curty}},
  \emph{\bibinfo{title}{Complete hierarchies of efficient approximations to
  problems in entanglement theory}}, \bibinfo{journal}{Physical Review A}
  \textbf{\bibinfo{volume}{70}}, \bibinfo{pages}{062317}
  (\bibinfo{year}{2005}), \eprint{quant-ph/0407135}.

\bibitem[{\citenamefont{Labuschagne et~al.}(2003)\citenamefont{Labuschagne,
  Majewski, and Marciniak}}]{LMM03}
\bibinfo{author}{\bibfnamefont{L.}~\bibnamefont{Labuschagne}},
  \bibinfo{author}{\bibfnamefont{W.}~\bibnamefont{Majewski}}, \bibnamefont{and}
  \bibinfo{author}{\bibfnamefont{M.}~\bibnamefont{Marciniak}},
  \emph{\bibinfo{title}{On {$k$}-decomposability of positive maps}}
  (\bibinfo{year}{2003}), \eprint{math-ph/0306017}.

\end{thebibliography}

\end{document}